\documentclass[mathleft]{an}
\usepackage{graphicx}
\usepackage{times}
\usepackage[font=small, labelfont=bf]{caption}
\def\gx{GX\,339--4}
\def\ctss{\,cts\,s$^{-1}$}
\def\phcms{\,ph\,cm$^{-2}$\,s$^{-1}$}
\def\deg{$^\circ$}
\def\rxte{\textit{RXTE}}
\def\swift{\textit{Swift}}
\begin{document}

\Pagespan{1}{}
\Yearpublication{2016}%
\Yearsubmission{2015}%
\Month{11}%
\Volume{---}%
\Issue{---}%
\DOI{---}

\title{Systematic spectral analysis of \gx: influence of Galactic background and reflection models}

\author{M.\ Clavel\inst{1}\fnmsep\thanks{Corresponding author:
  \email{maica.clavel@ssl.berkeley.edu}\newline}
\and  J. Rodriguez\inst{1}
\and S. Corbel\inst{1}
\and M. Coriat\inst{2,3}
}
\titlerunning{Spectral analysis of \gx}
\authorrunning{M.\ Clavel, et al.}
\institute{
Laboratoire AIM, UMR 7158, CEA/CNRS/Univ. Paris Diderot, CEA DSM/IRFU/SAp, F-91191 Gif-sur-Yvette, France
\and 
Universit\'e de Toulouse, UPS-OMP, IRAP, Toulouse, France
\and
CNRS, IRAP, 9 av. Colonel Roche, BP 44346, F-31028 Toulouse cedex 4, France
}

\received{7 Sep 2015}
\accepted{21 Oct 2015}
\publonline{later}


\abstract{Black hole X-ray binaries display large outbursts, during which their properties are strongly variable. We develop a systematic spectral analysis of the 3--40\,keV \rxte/PCA data in order to study the evolution of these systems and apply it to \gx. Using the low count rate observations, we provide a precise model of the Galactic background at \gx's location and discuss its possible impact on the source spectral parameters. At higher fluxes, the use of a Gaussian line to model the reflection component can lead to the detection of a high-temperature disk, in particular in the high-hard state. We demonstrate that this component is an artifact arising from an incomplete modeling of the reflection spectrum.
  }

\maketitle

\section{Introduction}
\label{sec:intro}
Along the course of their large outbursts, the spectral shape of black hole X-ray binaries (BHXB) varies, tracing the evolution of a jet or a corona, of an accretion disk and of the associated reflection processes. 
How these physical structures form and evolve over time is still under investigation (e.g.\ Remillard \& McClintock 2006). To better constrain the theoretical models, one needs to compare the simulation outputs with the real observations, applying the same methods to both data sets. Therefore, we are developing a systematic procedure to reduce the \rxte/PCA data (Sec.\,\ref{sec:obs}) and to perform the corresponding spectral analysis (Sec.\,\ref{sec:spec}). In order to test our method and to obtain an initial set of generic spectral properties for BHXB, we use \gx\ which undergoes frequent outbursts, as prime example (e.g.\ Dunn et al.\ 2008). We present a first overview of the results (Sec.\,\ref{sec:results}) as well as two further studies aiming to improve our systematic procedure: a detailed modeling of the Galactic background at \gx's position (Sec.\,\ref{sec:bkg}) and a comparison of the spectral parameters obtained using different models for the reflection component (Sec.\,\ref{sec:twokeV}). 

\section{Observations and data reduction}
\label{sec:obs}
\gx's recurrent outbursts have been monitored with a large number of observatories at all wavelengths. In this work we primarily analyze the 3--40\,keV \rxte/PCA data in order to perform a systematic analysis of the spectral evolution of the source over several outbursts. To test the validity of our models we also extend the spectral energy band of our study to lower X-ray energies, making use of \swift\ data (0.6--8\,keV) obtained in quasi-simultaneity with a given \rxte\ observation. 

\subsection{\rxte/PCA}
\label{sec:obsRXTE}
The \rxte\ mission operated from December 1995 to January 2012, providing a quasi-systematic follow-up of the X-ray binary outbursts over 16 years. We reduced all the 1389 Proportional Counter Array (PCA) observations available for \gx\footnote{This corresponds to all \rxte/PCA observations pointing within one degree from \gx, ignoring slew and raster observations, as well as observations containing data gaps possibly affecting the data (obsIDs ending with G, T or U).}. We used the \texttt{HEASOFT} software suite v6.16 to reduce the corresponding data in a standard way, restricting it to the top layer of Proportional Counter Unit~2 (PCU2)\footnote{PCU2 is the only unit that was active during all \rxte\ observations, and we tested that, for our study, the spectra were not significantly improved when adding the photon counts from its second and third layers.}. We time-filtered the data, using \textit{maketime} and \textit{xtefilt}, to remove PCU2 breakdowns and to restrict it to the times when the elevation angle above the Earth is greater than 10\deg, and when the satellite pointing offset is less than 0.02\deg. 
The PCA response file was computed using \textit{pcarsp} and the instrumental noise was estimated using \textit{pcabackest} with the background model pca\_bkgd\_cmbrightvle\_eMv20051128.mdl. All observation/background spectra and average count rates were then extracted using \textit{saextrct}. For PCA observations having a net average count rate lower than 40\ctss\ over the full energy range, the instrumental background estimation was replaced by the one calculated from model pca\_bkgd\_cmfaintl7\_eMv20051128.mdl. All spectra were binned in order to have at least 200 counts per resultants channel and a systematic error of 0.6\% was added to account for instrumental uncertainties. In this work we present 3--10\,keV lightcurves and spectral fits obtained in the \hbox{3--40\,keV} energy range. The astrophysical background included in these data is investigated in Section~\ref{sec:bkg}.

\subsection{\swift/XRT}
\label{sec:obsSwift}
We reduced and analyzed one observation from \swift, corresponding to MJD\,55260.2, using the {\texttt{HEASOFT}} software suite v6.16. We considered only the data obtained with the X-ray Telescope (XRT) that covers the soft X-rays (0.5--10\,keV). This observation was obtained in window timing ({\texttt{wt}}) mode. The \swift/XRT level 2 cleaned event file was obtained with \textit{xrtpipeline}, and further processed within {\texttt{XSELECT}} to obtain the source and background spectra. As recommended in Evans et al.\ (2009) for the {\texttt{wt}} data, the source spectrum was extracted from a 30 pixel radius circular region centered on the best source position. The resultant data is not contaminated by pile-up effects. \\
\indent
The background spectrum was estimated from larger regions at off-axis positions. The ancillary response file (arf) was estimated with \textit{xrtmkarf}, and the last version (swxwt0to2s6\_20090101v015.rmf) of the redistribution matrix files (rmf) used in the spectral fits. The spectral channels were grouped in order to have at least 100\ctss\ per resultant channels, and only the data between 0.6--8\,keV were considered for the spectral fits.

\section{Spectral analysis}
\label{sec:spec}
Following the method proposed by Dunn et al.\ (2010), we perform a systematic analysis of the 1385 spectra\footnote{Four \rxte/PCA observations (MJD\,52095.999, 52128.566, 54327.758 and 54327.823) out of the 1389 we reduced had to be excluded due to a too low number of counts and/or data reduction issues preventing a full spectral analysis.}. For each observation, we test the presence of several spectral components using {\texttt XSPEC} software v12.8.2 models and chi-squared fitting routines, as well as parameter constraints discussed by Dunn et al.\ (2008) and Plant et al.\ (2014). 

\subsection{Choice of model sets}
In addition to the non-thermal component modeled by an absorbed power law, {\sc phabs$\times$powerlaw}\footnote{A model including a broken power law and/or a high energy cut-off, as tested by Dunn et al. (2010), are not statistically needed within the energy range of our analysis (3--40\,keV).}, we test the presence of a thermal emission coming from the accretion disk modeled by a multi-temperature blackbody {\sc ezdiskbb}, and of a reflection component.\\
\indent
The model chosen by Dunn et al.\ (2010) to account for the reflection component is a Gaussian emission line {\sc gauss} fixed at ${\rm E_{line}}=6.4$\,keV, only modeling the strong iron fluorescent line. From our analysis, it appears that a smeared absorption edge at ${\rm E_{edge}}=7.1$\,keV, corresponding to the photoelectric absorption by iron atoms, is also significant in part of the observations (Sec.\,\ref{sec:twokeV}). Therefore, we perform a second systematic fitting, including the multiplicative {\sc smedge} model when the Gaussian line is present. Finally, we run a third round of fits testing the self consistent model  {\sc xillver} to account for the reflection component. This model includes the photoelectric absorption, the fluorescent emission lines and the Compton scattering continuum emission (Garc\'ia et al.\ 2013).
\\
\indent
The three sets of {\texttt XSPEC} models that we test are summarized by the following three equations:
\begin{small}
\begin{equation}
	{\rm \sc{phabs}}\times({\rm \sc{powerlaw}}+{\rm \it{ezdiskbb}}+{\rm \sc{\bf{gauss}}})
	\label{eq:1}
\end{equation}
\begin{equation}
	{\rm \sc{phabs}}\times({\rm \sc{powerlaw}}+{\rm \it{ezdiskbb}}+{\rm \sc{\bf{gauss}}})\times{\rm \sc{\bf{smedge}}}
	\label{eq:2}
\end{equation}
\begin{equation}
	{\rm \sc{phabs}}\times({\rm \sc{powerlaw}}+{\rm \it{ezdiskbb}}+{\rm \sc{\bf{xillver}}})
	\label{eq:3}
\end{equation}
\end{small}
where the disk component (in italic) and of the reflection component (in bold) are included only if they are statistically significant (see Sec.\,\ref{sec:ftest} for the model selection procedure). The column density corresponding to the photoelectric absorption {\sc phabs} is set to ${\rm N_H}=0.4\times10^{22}$\,cm$^{-2}$. The Gaussian line energy is fixed to ${\rm E_{line}}=6.4$\,keV, the {\sc smedge} threshold energy to ${\rm E_{edge}}=7.1$\,keV, its smearing width to ${\rm W}=15$\,keV and the index for photoelectric cross section to ${\rm a}=-2.67$. The {\sc xillver} incident radiation is fixed to the {\sc powerlaw} parameters, the iron abundance to the solar value, and the system inclination to $i=40$\deg. All the other parameters (power law photon index $\Gamma$ and normalization ${\rm I_{pwl}}$, disk maximal temperature ${\rm T_{max}}$ and normalization ${\rm I_{disk}}$, Gaussian normalization $\rm I_{line}$, edge maximum absorption factor at energy threshold $\rm \tau_{max}$, {\sc xillver} ionization parameter $\xi$ and normalization $\rm I_{refl}$) are left free to vary within physical ranges and are initialized at typical values in order to increase the chance of converging rapidly to a consistent fit.

\subsection{Procedure to select the best fit model}
\label{sec:ftest}
For a given reflection model, our systematic analysis perform the spectral fit of each observation with four models ({\sc powerlaw}, {\sc powerlaw+reflection}, {\sc powerlaw+disk} and {\sc powerlaw+disk+reflection}). We then identify the model having the best chi-square and use the {\texttt XSPEC} F-statistic test to check the statistical relevance of any optional components (ie. any disk and/or reflection components), based on a comparison of chi-squares ($\chi^2$) and degrees of freedom (dof).\\
\indent
We are aware that the F-test we use is limited to data sets having a Gaussian statistic (which is the case considering the data large count rates and the consequent binning applied to the spectra, see Sec.\,\ref{sec:obsRXTE}) and cannot be used to test the presence of a Gaussian line (Protassov et al. 2002) nor of a multiplicative component (Orlandini et al. 2012). Therefore, similarly to what is done by Dunn et al. (2010) we added a test on the iron line parameters: it is considered statistically relevant only if its normalization is at least 2\,$\sigma$ above zero. When the emission line is detected, the edge should also be present in the spectrum. However, this multiplicative component may not be statistically needed for the fit. This is why we decide to include the {\sc smedge} in the best fit model only if the following two conditions are fulfilled: (i) the Gaussian line is required, (ii) the edge is significantly improving the overall fit (lowering the reduced $\chi^2$ from more than 2.5 down to about~1).\\
\indent
The phenomenological models we are testing are quite simple. Therefore, for most of the observations, the best fit model selection can easily be done by eye. When this is the case, the automatic procedure gives results that are in good agreement with what one would expect (except for model set\,\ref{eq:1}, which is further discussed in Sec.\,\ref{sec:twokeV}). For this reason, we decide not to investigate further the statistical tests. Apart from few isolated observations (less than 1\% of our sample), all best fit models have reduced $\chi^2$ around 1 and the uncertainties we provide correspond to the 1\,$\sigma$ error~bars.

\section{Standard spectral evolution of \gx}
\label{sec:results}
\begin{figure}[ht]
	\centering
	\includegraphics[trim = 0mm 0mm 0mm 0mm, clip, width=\columnwidth]{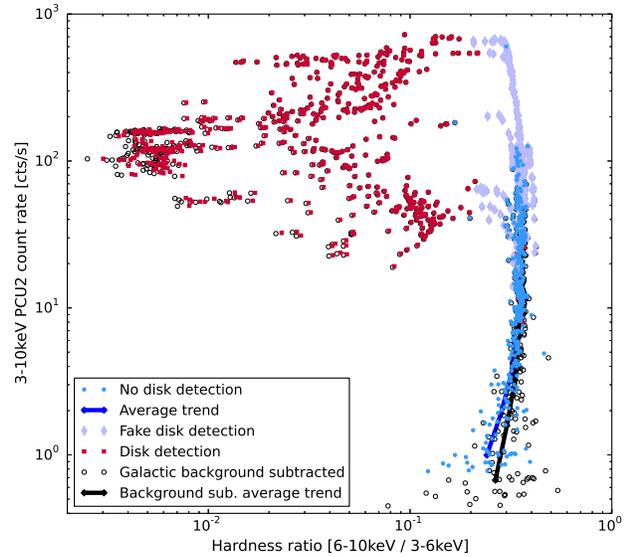}
	\vspace{-0.4cm}
	\caption{Hardness intensity diagram for all \rxte\ observations having a net average count rate above 0.77\ctss. The intensity is given as a number of counts detected in PCU2 top layer between 3 and 10\,keV, while the hardness ratio is defined as the ratio of the 6--10\,keV flux over the 3--6\,keV one. The different colors represent the spectral shape of the data with a discrimination whether a disk component is statistically needed (red squares) or not (blue circles). The purple diamonds highlight a region where a spurious disk can be detected, depending on the model chosen to account for the reflection component (see also Fig.\,\ref{fig:LC} and Sec.\,\ref{sec:twokeV}). The black empty circles are the same observations corrected from the Galactic background component (Sec.\,\ref{sec:bkg}). The softening visible at low count rates (average trend shown by the blue line) is partly removed when excluding the Galactic background (black line).}
	\label{fig:HID}
\end{figure}

\begin{figure*}[ht]
	\centering
	\includegraphics[trim = 0mm 0mm 0mm 2mm, clip,width=0.95\textwidth]{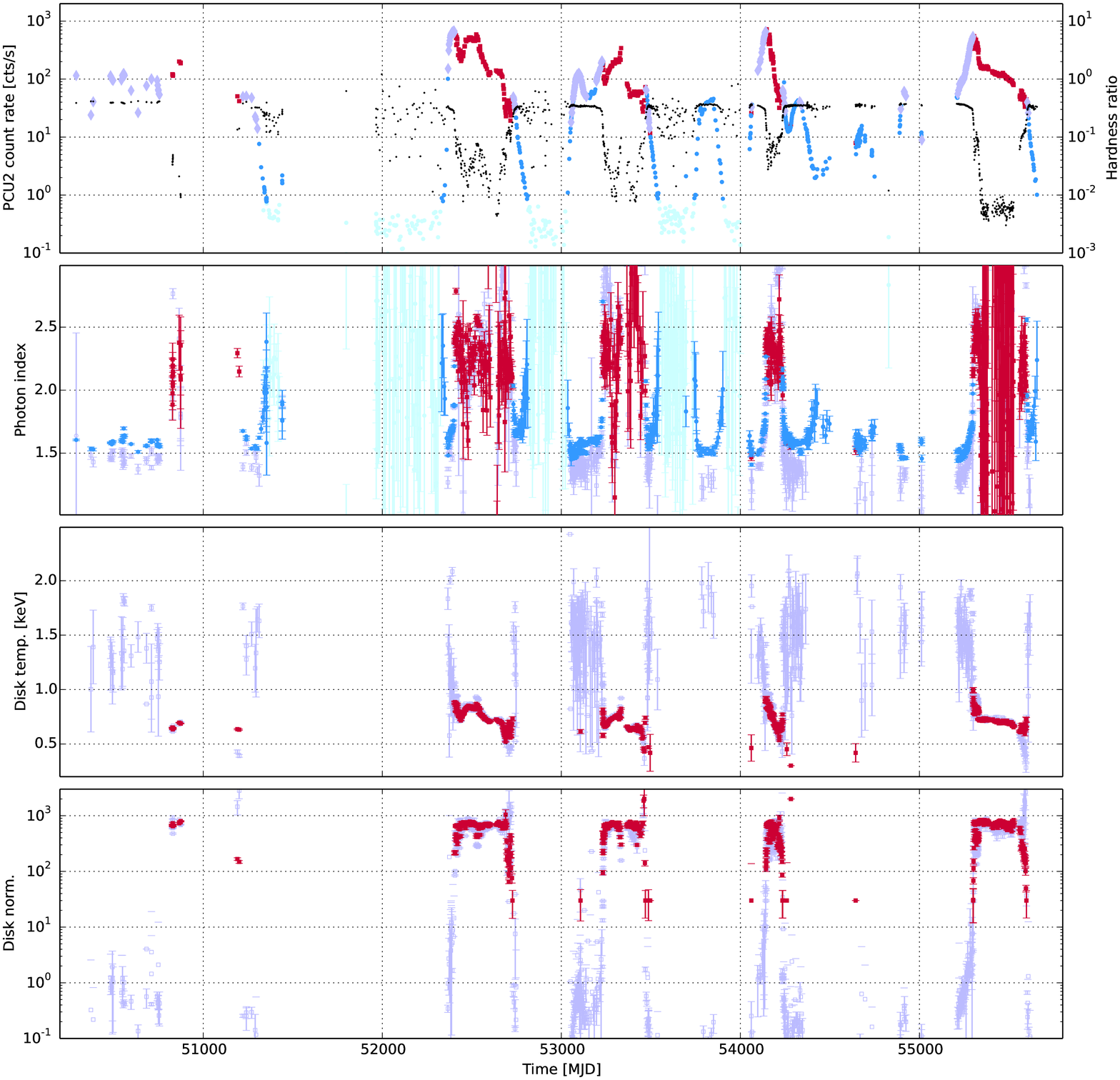}
	\hspace{-1cm}
	\includegraphics[trim = -30mm 0mm 0mm 0mm, clip, scale=0.55, angle=90]{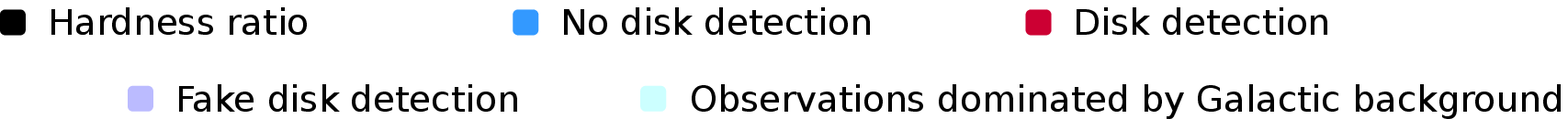}
	\caption{\rxte/PCA lightcurve of \gx\ and corresponding spectral components. From top to bottom: (i) 3--10\,keV lightcurve with a color coding highlighting what observations are modeled with a simple absorbed power law (blue circles) and which ones need a disk component (red squares). The purple diamonds highlight periods where model set (\ref{eq:1}) gives a best model including a 1.5\,keV disk while model sets (\ref{eq:2}) and (\ref{eq:3}) do not statistically require an additional disk component (Sec.\,\ref{sec:twokeV}). Data points in lightblue have net count rates below 0.77\ctss\ and are considered as dominated by the Galactic background (Sec.\,\ref{sec:bkg}). The hardness ratio of each observation (black dots) is defined in the caption of Fig.\,\ref{fig:HID}; (ii) Power law photon index of the best fit obtained using model set (\ref{eq:1}, purple) or (\ref{eq:3}, same color coding as top panel). The spurious detection of the 1.5\,keV disk is responsible for an artificially low photon index in the corresponding periods; (iii) Maximum temperature of the disk found using model set (\ref{eq:1}, purple) or (\ref{eq:3}, red); (iv) Normalization of the disk (same color coding). The fake disk component appears in purple with a high temperature and a very low normalization, while for the true detections, red and purple points are overlaid (see Sec.\,\ref{sec:twokeV} for more details).}
	\label{fig:LC}
\end{figure*}

Our data set is made of 1385 observations spread over the entire life time of the \rxte\ mission, sampling five major outbursts of \gx\ (see the source lightcurve in Fig.\,\ref{fig:LC}). These events all follow a standard cycle in the Hardness Intensity Diagram (HID, Fig.\,\ref{fig:HID}) that can be summarized as follow: (i) an intensity rise in the hard state, best modeled by an absorbed power law, (ii) a transition to the soft state, with the detection of an additional disk component, (iii) an intensity decrease in the soft state, where the disk component is dominating the spectra and the power law component is often poorly constrained, (iv) a transition back to the hard state, with the disappearance of the disk, and (v) an intensity decrease with a source remaining in the hard state.\\
\indent
If a sufficient model is used for the reflection component (i.e. model set \ref{eq:2} or \ref{eq:3}, see further discussion in Sec.\,\ref{sec:twokeV}), we can identify typical parameters following the spectral evolution of the source (Fig.\,\ref{fig:LC}, middle panels). The power law photon index goes from about 1.6 in the hard state to about 2.3 during the transition, and it is poorly constrained both during the softest observations and the lowest count rate ones (dominated by the Galactic background emission, Sec.\,\ref{sec:bkg}). The disk temperature varies from about 1\,keV for the brightest down to about 0.5\,keV for the faintest observations in the soft state. These results are in agreement with previous works (Dunn et al. 2008, Plant et al. 2014).

\section{Galactic background}
\label{sec:bkg}
\gx\ is located within the Norma arm, in the Galactic plane. Therefore, a diffuse emission from the Galactic background, also called the ridge emission, should be present in the region (Valinia \& Marshall 1998). Assuming that this astrophysical background emission is constant over the 16~years of \rxte\ observations, we provide a precise estimation of this component at the position of \gx\ and investigate the consequences of neglecting it, regarding the best fit model parameters.

\subsection{Galactic background estimation from \rxte\ data}
\label{sec:bkgparam}
Our set of data includes a large number of low count rate observations, allowing for a precise estimation of the contribution of the Galactic background detected by PCA at the position of \gx. Indeed, the histogram giving the number of observations as a function of their average net count rate in the 3--10\,keV range can be described as a Gaussian distribution centered on 0.304\ctss\ (with a standard deviation $\sigma=0.076$\ctss), plus individual observations ranging from 0.77\ctss\ (3\,$\sigma$) up to about 800\ctss. According to previous works, the distribution at low fluxes is dominated by the Galactic background component (e.g. Coriat et al. 2009). We select the 125 observations having a count rate within 2\,$\sigma$ of the  Gaussian distribution mean and add them is order to get a spectral estimation of the background component with a sufficient exposure to perform a spectral fit  (about 140\,ks total). \\
\indent
Valinia \& Marshall (1998) estimated the ridge emission from \rxte\ observations covering the central portion of the Galactic disk ($-45^\circ<{\rm l}<+45^\circ$, $-1.5^\circ<{\rm b}<+1.5^\circ$), and provided an average model for this large region: 
\begin{equation}
	{\rm wabs}\times({\rm raymond} + {\rm powerlaw})
	\label{eq:bkg}
\end{equation}
where {\sc raymond} is a model for a hot and diffuse gas at redshift $z=0$, with a solar abundance, a temperature of about ${\rm kT}\sim2.9$\,keV and a normalization ${\rm I_{ray}}\sim0.021$\,cm$^{-5}$. The power law photon index is $\Gamma\sim1.8$, and its normalization $\rm I_{pwl}\sim0.00391$\phcms\,keV$^{-1}$ at 1\,keV. The overall absorption is ${\rm N_H}\sim1.8\times10^{22}$\,cm$^ {-2}$. The average Galactic background estimated by these authors, corresponds to a flux  ${\rm F_{ridge}}\sim2.9\times10^{-3}$\phcms\ in the 3--40\,keV range. \\
\indent
We fit this model to the astrophysical background at the position of \gx, fixing all the parameters to Valinia \& Marshall (1998) values except for the normalizations of the plasma and power law components. The fit is satisfactory ($\chi^2$ / dof $=$ 59 / 64) and the parameters found are ${\rm I_{ray}}=(4.4\pm0.6)\times10^{-3}$\,cm$^{-5}$ and $\rm I_{pwl}=(7.8\pm1.2)\times10^{-4}$\phcms\,keV$^{-1}$, which are five times lower than the average values obtained by Valinia \& Marshall (1998), and correspond to a 3--40\,keV flux ${\rm F_{bkg}}\sim5.9\times10^{-4}$\phcms. Since the ridge emission is not supposed to be uniform along the Galactic plane, we believe that the best fit model we obtained for our average low count rate spectrum is relevant to model the Galactic background at \gx's position.

\subsection{Neglecting the Galactic background component}
The estimated Galactic background has a rather soft spectrum (equivalent to a photon index $\Gamma\sim2.2$, if fitted with an absorbed power law\footnote{The fit is performed fixing the absorption ${\rm N_H}$ to \gx\ standard value and leaving the power law photon index and normalization free. This is not the best fit model for the average spectrum ($\chi^2$ / dof $=$ 100 / 64). However, this model is sufficient for individual observations having a lower statistic (Fig.\,\ref{fig:LC}).}) and is likely to affect the low-hard state observations if neglected. In order to quantify this effect, we compare systematic fits performed using model set~(\ref{eq:2}) to the same set including an additional Galactic background component (given by equation \ref{eq:bkg} and fixed to the parameters derived in Sec.\,\ref{sec:bkgparam}). All observations dominated by the Galactic background (net count rates below 0.77\ctss\ in the 3--10\,keV range) are poorly fitted. For all observations dominated by \gx, the two model sets give consistent results concerning the spectral components being statistically significant, and the spectral model parameters are compatible within the error bars. Therefore we decide to restrict the study of the spectral evolution of \gx\ to the observations above 0.77\ctss, and to neglect the Galactic background contribution for these observations.\\
\indent
However, part of the softening trend visible for the lower fluxes (increasing photon index towards low count rates, Fig.\,\ref{fig:LC}) is partly corrected when including the Galactic background as a fixed component in the spectral fit, as shown by the corresponding average trends in Fig.\,\ref{fig:HID}. From our analysis we believe that the softening effect only becomes truly negligible above about 8\ctss\ (in the 3--10\,keV range). Therefore the average spectral parameters obtained from the faintest observations should be interpreted with caution. 

\section{Importance of the reflection model}
\label{sec:twokeV}
We perform the whole systematic fitting procedure three times, using three different models to account for the reflection component (Sec.\,\ref{sec:spec}). With the \rxte/PCA data alone we do not have the spectral resolution to extract precise information about the reflection phenomena, therefore we cannot provide a precise comparison between the three reflection models. However, model set~(\ref{eq:1}), including only the fluorescent line at 6.4\,keV, leads to the detection of a high temperature disk that is not statistically needed when choosing more complete models for the reflection component (Fig.\,\ref{fig:LC}, bottom panels).

\subsection{Spurious detection of a $\sim$1.5\,keV disk}
\begin{figure}[ht]
	\centering
	\includegraphics[trim=0mm 0mm 25mm 5mm, clip,width=\columnwidth]{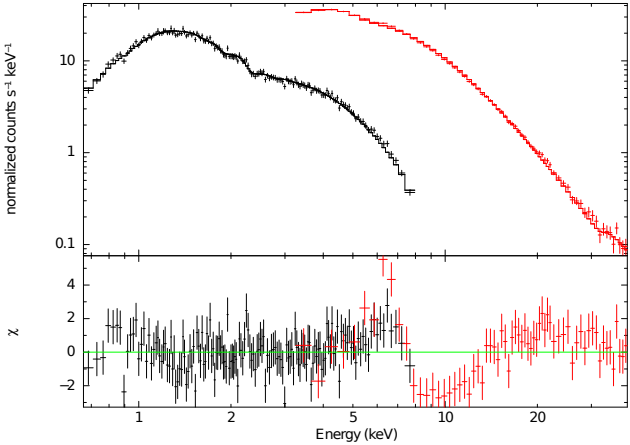}
	\caption{Simultaneous fit of the \swift\ MJD\,55260.2 (black) and \rxte/PCA MJD\,55260.4 (red) spectra of \gx\ with an absorbed power law. The residuals computed from this poor fit ($\chi^2$~/~dof $=$ 340 / 213) highlight the presence of reflection features between 5 and 20\,keV.}
	\label{fig:swift}
\end{figure}

For all \gx\ outbursts, when using model set (\ref{eq:1}) we significantly detect a disk component in the high-hard state observations (see also Dunn et al.\ 2008, Nandi et al.\ 2012). This disk is best described by a high maximal temperature ${\rm kT_{max}} \sim 1.5$\,keV and a low normalization ${\rm I_{disk}}<10$, it is also associated to a harder power law than the typical values found with the other two model sets (Fig.\,\ref{fig:LC}). We believe that the high-temperature disk is an artifact created by the spectral analysis, for three main reasons.\\
\indent
First, the normalization of the disk is defined by:
\begin{equation}
	{\rm I_{disk}} = \frac{1}{f^4}\left(\frac{R_{in}}{1 {\rm km}}\right)^2\left(\frac{D}{10 {\rm kpc}}\right)^{-2} \cos i 
	\label{eq:disknorm}
\end{equation}  
where, for \gx, $f = 1.7$ is the color effective ratio, $D\sim8$\,kpc is the distance of the source, $i\sim40^\circ$ is the system inclination and $R_{in}$ is the inner radius of the disk, which should be larger than the Schwarzschild radius ${\rm R_S} \sim 30$\,km for a 10\,M$_\odot$ black hole. Therefore, for \gx\ we expect the disk normalization to be not less than about 120 (when $R_{in} = {\rm R_S}$) or 30 to be conservative. With such a low normalization, the high-temperature disk parameters are not consistent with the complete disk model we are using.\\ 
\indent
Second, even if the {\sc ezdiskbb} model does not seem relevant, the high-temperature disk could fit a low-energy excess in the spectra (e.g.\ thermally emitting disk fragments that are embedded in a hot and optically thin disk). However, when available the simultaneous \swift/XRT data do not highlight any need for an additional thermal component at low energy. The \swift\ observation is well fitted by an absorbed power law having the following parameters: ${\rm N_H}=(0.37\pm0.01)\times10^{22}$\,cm$^{-2}$, $\Gamma = 1.53\pm0.02$ and ${\rm I_{pwl}} = 0.39\pm0.01$\phcms\,keV$^{-1}$ ($\chi^2$ / dof = 147 / 148). When added to the fit, the quasi-simultaneous \rxte/PCA data show discrepancies compared to the single power law model. The residuals highlight a possible emission line and associated absorption edge within the 5--20\,keV range  (Fig.\,\ref{fig:swift}). In order to avoid any strong biased linked to the reflection component, we perform a simultaneous fit ignoring data points between 5 and 20 keV for both instruments. The fit is satisfactory and the parameters found are compatible with the one listed previously, within the 1\,$\sigma$ error bars ($\chi^2$ / dof = 166 / 176), confirming that the high-energy power law is fully compatible with the low-energy one. Therefore, there is no need for any thermal component at any temperature. Simultaneous fits using model set (\ref{eq:2}) or (\ref{eq:3}) on the whole 0.6--40\,keV energy range are both satisfactory with no disk component needed ($\chi^2$ / dof $\sim$ 1).\\
\begin{table*}[ht]
\centering
\caption{Models (A) and (B) are statistically equivalent for \rxte/PCA observations within 3--40\,keV (parameters as defined in Sec.\,\ref{sec:spec}).}
\label{tab:fakeit}
\begin{tabular}{l c c c c c c c c c c}\hline
Parameters & $\Gamma$ & ${\rm I_{pwl}}$ & $\rm E_{line}$ & $\sigma$ & ${\rm I_{line}}$ & $\rm E_{edge}$ & $\rm \tau_{max}$& $\rm W$ & $\rm kT_{max}$ & $\rm I_{disk}$\\
 &  & (\phcms\,keV$^{-1}$) & (keV) &  (keV) & ($10^{-4}$\phcms) & (keV) & & (keV) & (keV) & \\
\hline
(A) & 1.5 & 0.20 & 6.4 & 0.1 & 4 & 7.1 & 0.8 & 15 & -- & --\\
(B) & 1.4 & 0.15 & 6.4 & 0.8 & 0.15 & -- & -- & -- & 1.44 & 0.38\\
\hline
\end{tabular}
\end{table*}
\indent
Third, to quantify the influence on the best fit parameters of not fitting the absorption edge, we created a fake spectrum using the following model: {\sc phabs $\times$ (powerlaw + gauss) $\times$ smedge} and run the fitting procedure using model set~(\ref{eq:1}). The fit is satisfactory ($\chi^2$ / dof = 50 / 59), the initial and output model parameters are summarized in Table \ref{tab:fakeit}. The best fit has a harder power law than the initial one, better fitting the absorption edge, and resulting in an artificial excess of emission at low energy that is fitted by a high-temperature disk with a low normalization. This is exactly what we are witnessing when using model set (\ref{eq:1}) on the \rxte/PCA data.
Therefore, we believe that the high-temperature disk component is not physical but arises from an incomplete modeling of the reflection spectrum.

\subsection{Converging to a coherent solution}
As demonstrated, the spurious detection of the high-temperature disk is due to a limited spectral range, an incomplete model set and a disk normalization not bound to physical values. In order to exclude this fake solution within our systematic analysis, we first add a hard lower limit $\rm I_{disk} > 30$ for the disk normalization parameter\footnote{This lower limit has been tested for \gx's \rxte\ spectra only. It may not be relevant to other BHXB with more complex disk geometries.}. This alone excludes the high-temperature disk solution but does not provide a satisfactory fit for the corresponding observations. Second, we test the presence of the absorption edge, whenever the line is present. This multiplicative component is sufficient to provide a satisfactory fit for all observations from which we detected a spurious high-temperature disk. Third, in order to verify that this supplementary component is linked to the reflection spectrum we compare the phenomenological model set (\ref{eq:2}) to model set (\ref{eq:3}), which includes a self-consistent model for the reflection features. These two analyses provide consistent results regarding the fit parameters, validating the use of an absorption edge when statistically needed.

\section{Conclusion}
Our systematic analysis of the \rxte/PCA data between 3--40\,keV highlight similar trends for all \gx\ outbursts, and it is important to disentangle the variations tracing the true evolution of the source from the one generated by the analysis itself. In this work we demonstrated that neglecting the Galactic background is justified but leads to an artificial softening of the source spectra for the lowest count rate observations. At higher fluxes, a limited energy range and a possibly incomplete model set can result in the detection of spurious components, such as the high-temperature disk which artificially substitutes for missing features in the reflection model. Therefore, physical inputs on the source, as well as multi-wavelength observations, are crucial to fully test our systematic analysis and to provide the generic properties of \gx. This extensive study is beyond the scope of this paper and will be presented in a future publication (Clavel et al.\ in prep).


\acknowledgements
The authors acknowledge funding support from the French Research National Agency: CHAOS project ANR-12-BS05-0009 (http://www.chaos-project.fr).



\begin{thebibliography}{}
\vspace{-0.2cm}
\bibitem[\protect\citeauthoryear{Coriat et al.}{2009}]{coriat2009} Coriat M., Corbel S., Buxton M.~M., Bailyn C.~D., Tomsick J.~A., K{\"o}rding E., Kalemci E.: 2009, MNRAS 400, 123

\bibitem[\protect\citeauthoryear{Dunn et al.}{2008}]{dunn2008} Dunn R.~J.~H., Fender R.~P., K{\"o}rding E.~G., Cabanac C., Belloni T.: 
2008, MNRAS 387, 545 

\bibitem[\protect\citeauthoryear{Dunn et al.}{2010}]{dunn2010} Dunn R.~J.~H., Fender R.~P., K{\"o}rding E.~G., Belloni T., Cabanac C.: 2010, MNRAS 403, 61

\bibitem[\protect\citeauthoryear{Evans et al.}{2009}]{evans2009}{Evans}, P.~A., {Beardmore}, A.~P., {Page}, K.~L., et al.: 2009, MNRAS 397, 1177

\bibitem[\protect\citeauthoryear{Garc{\'{\i}}a et al.}{2013}]{Garcia2013} Garc{\'{\i}}a, J., Dauser, T., Reynolds, C.~S., {Kallman}, T.~R., {McClintock}, J.~E., {Wilms}, J., 	{Eikmann}, W.: 2013, ApJ 768, 146 

\bibitem[\protect\citeauthoryear{Nandi et al.}{2012}]{nandi2012} Nandi A., Debnath D., Mandal S., Chakrabarti S.~K., 2012, A\&A, 542, A56

\bibitem[\protect\citeauthoryear{Orlandini et al.}{2012}]{orlandini2012} Orlandini M., Frontera F., Masetti N., Sguera V., Sidoli L., 2012, ApJ, 748, 86 

\bibitem[\protect\citeauthoryear{Plant et al.}{2014}]{plant2014} Plant D.~S., Fender R.~P., Ponti G., Mu{\~n}oz-Darias T., Coriat M.: 2014, MNRAS 442, 1767 

\bibitem[\protect\citeauthoryear{Protassov et al.}{2002}]{protassov2002} Protassov R., van Dyk D.~A., Connors A., Kashyap V.~L., Siemiginowska A., 2002, ApJ, 571, 545 

\bibitem[\protect\citeauthoryear{Remillard \& McClintock}{2006}]{remillard2006} Remillard R.~A., McClintock J.~E.: 2006, ARA\&A 44, 49

\bibitem[\protect\citeauthoryear{Valinia \& Marshall}{1998}]{valinia1998} Valinia A., Marshall F.~E.: 1998, ApJ 505, 134


\end{thebibliography}
\end{document}